\documentclass{llncs}
\usepackage[english]{babel}

\usepackage{graphicx}

\begin{document}
\title{Simulation of Computer Network Attacks} 
\author{
Carlos Sarraute \inst{1,2}
\and Fernando Miranda \inst{1} 
\and Jose I. Orlicki \inst{1,2} 
}
\institute{
CoreLabs, Core Security Technologies
\and ITBA (Instituto Tecnol\'ogico de Buenos Aires)
}
\maketitle

\begin{abstract}

In this work we present a prototype for simulating computer network attacks.
Our objective is to simulate large networks 
(thousands of hosts, with applications and vulnerabilities)
while remaining realistic from the attacker's point of view.
The foundation for the simulator is a model of computer intrusions,
based on the analysis of real world attacks.
In particular we show how  to interpret vulnerabilities and exploits
as communication channels.
This conceptual model gives a tool to describe the theater of operations, 
targets, actions and assets involved in 
multistep network attacks.
We conclude with applications of the attack simulator.

\end{abstract}

\section{From Real World Attacks to an Attack Model} \label{attacks}

\subsection{Introduction}

In Section \ref{attacks} we begin with a brief description of real world attacks,
and describe an abstraction of the attack actions.
Section \ref{exploits} delves in mores detail in the ``Attack and Penetrate" phase of an attack,
in which the attacker exploits a  vulnerability to gain access to a target machine.
Again, we show how to abstract and generalize the process of exploitation and payload execution.
In Section \ref{simulator} we present the prototype for simulating network attacks,
based on the implementation of multiplatform agents
and the abstraction of vulnerabilities and exploits as communication channels.
Section \ref{performance} deals with the tension between realism and performance
in the simulation.
Finally we mention some applications of the simulator.

\subsection{Computer Network Intrusions}

During a network intrusion, an attacker tries to 
gain access to software systems that require authorization
(for example, web servers, database servers or accounting systems).
The intrusion may be illegal (what people usually have in mind when 
speaking about intrusions), or may be an authorized audit performed by security professionals.
The latter is called a network penetration test: a proactive and authorized attempt to compromise 
network security and access sensitive information by taking advantage of vulnerabilities.
As networks evolve, and combine a multitude of interconnected technologies, the penetration test
has become an accepted practice to evaluate the global security of a network
(ultimately assessing effectiveness of the deployed security countermeasures).
The interesting point for us is that pentesters basically use the same tools and methodologies
as unauthorized attackers, so we can focus on the former (whose practices are also more documented!)

\subsection{Main Steps of an Attack}
\label{main_steps}

\subsubsection{Information Gathering.}

A successful attack depends on the ability to gather relevant information about the
target network,
including active Internet Protocol (IP) addresses, operating systems and available services. 
Actions realized during this phase include:
\begin{itemize}
\item{
Network discovery: performed utilizing mechanisms including ARP, TCP SYN packets, ICMP echo request, 
TCP connect and passive discovery. }
\item{
Port scanning: an exhaustive scan of open and closed ports of all the network hosts.}
\item{
OS identification: consists in recognizing the OS of a remote host by analyzing its responses to a set of tests. 
Classical Nmap's fingerprinting database can be combined with a neural network to accurately 
match OS responses to signatures, see \cite{neural_network}.
Additional OS identification capabilities are available for more specific situations. 
For instance, OS detection utilizing the DCE-RPC and SMB protocols can identify Windows machines more precisely.}
\item{Other techniques available to human attackers are social engineering and 
Google hacking (using publicly available information to gain insight on the target organization).
}
\end{itemize}

\subsubsection{Attack and Penetrate.}
During this phase, the attacker selects and launches remote exploits
making use of data obtained in the Information Gathering step. 
An exploit is a piece of software that injects code in the vulnerable system's memory
and modifies  the execution flow to make the system run the exploit code.
As we will see in section \ref{exploits}, the exploit can be thought of
as a way to install an agent on a compromised host.

\subsubsection{Local Information Gathering.}
The Local Information Gathering step collects information about computers that the attacker
has successfully compromised. During this phase, 
the attacker may gather information about the OS, network configuration, users and installed applications;
browse the filesystem on compromised systems;
view rights obtained and interact with compromised systems via shells.

\subsubsection{Privilege Escalation.}
During the Privilege Escalation phase, the attacker attempts to penetrate deeper into a
compromised computer by running local exploits in an attempt to obtain administrative
privileges (to gain root or superuser privileges).

\subsubsection{Pivoting.}
After Privilege Escalation, the attacker can use the newly controlled host as a vantage point
from which to run attacks deeper into the network.
By sending instructions to an installed agent, the attacker
can run local exploits to attack systems internally, rather than from across the network.
He can view the networks to which a compromised computer is connected,
and launch attacks from any compromised system to other computers on the same network,
gaining access to systems with increasing levels of security.
That is, the attacker executes the previous steps (Information Gathering and Attacking)
using the new agent as a source.

\subsubsection{Clean Up.}
The attackers needs to clean up his steps to avoid detection. Towards this end, all the executed actions
should minimize the noise produced, for example, by making modifications only in memory
and by not writing files in the target's filesystem.

\subsection{Abstraction of Attack Actions}

After this brief review of the steps of a real world attack, we present here the model 
that we use as abstraction of an attack.
The conceptual building blocks are Assets, Actions and Agents.

\subsubsection{Assets.}

An asset can represent anything that an attacker may need to
obtain during the course of an attack. More precisely, it 
represents the knowledge that an attacker has of a real object 
or property of the network. Examples of assets are: \\
{\tt 
\indent * BannerAsset (banner, host, port) \\
\indent * OperatingSystemAsset (os, host) \\
\indent * IPConnectivityAsset (source, target) \\
\indent * TCPConnectivityAsset (source, target, port) }

\noindent
A BannerAsset represents the {\tt banner} that an attacker
obtains when trying to connect to a certain {\tt port} on a {\tt host}. 
An OperatingSystemAsset represents the knowledge that an attacker
has about the operating system of a {\tt host}.
An IPConnectivityAsset represents the fact that an attacker is able to establish
an IP connection between a source host and a target host (given by their IP addresses).
A TCPConnectivityAsset represents the fact that an attacker is able
to establish a TCP connection between a {\tt source} host and a fixed
{\tt port} of a {\tt target} host (given by port number and IP address).

The assets we consider are probabilistic. This allows us to represent
properties which we guess are true with a certain probability
or negative properties (which we know to be false). For example,
an action which determines the OS of a host using banners 
(OSDetectByBannerGrabber) may give as result an 
OperatingSystemAsset {\tt os=linux} with {\tt probability=0.8} and
a second one with {\tt os=openbsd} and {\tt probability=0.2}. 
Another example, an ApplicationAsset {\tt host=192.168.13.1}
and {\tt application=\#Apache} with {\tt probability=0} means that 
our agent has determined that this host is not running Apache.

\subsubsection{Attack Actions.} \label{actions}

These are the basic steps which form an attack.
Examples of actions include, but are not limited to:
Apache Chunked Encoding Exploit, WuFTP globbing Exploit
(subclasses of Exploit), Banner Graber, OS Detect by Banner,
OS Fingerprint, Network Discovery, IP Connect and TCP Connect.
We review below the principal attributes of an action.

\subsubsection{Action goal.}
An action has a goal and when executed successfully the action completes the asset
associated with its goal. This is also called the \emph{action result}.

Usually, an action is directed against a target,
where the target is a computer or a network.
But there are different types of goals like gathering information
or establishing connectivity between two agents or hosts,
where the notion of target is not so clear. Thus the concept
of goal is more general and allows us to speak about intermediate
steps of an attack.

It is also common to speak about the result of an action
(for example to increase access, obtain information, corrupt information, 
gain use of resources, denial of service), focusing on non authorized results.
This is a particular case of our concept of goal.
Note that when an action completes the goal asset,
we are taking into account only the expected result of the action.
Undesired results and other side effects fall into the category of noise.

\subsubsection{Action requirements.}
The {\em requirements} are assets that will be the goals of other attack actions, 
which must have been successfully
executed before the considered action can be run. 
The requirements are the equivalent of children nodes in \cite{Sc00}
and subgoals in \cite{TLK01} and \cite{MEL01}.
An abstract action must know what kind of assets it may satisfy
and which goals it requires before running. These relations can be
used to construct an attack graph. 
By analyzing of the attack graph, the attacker can build a plan 
(as a sequence of actions) to reach the final objective.
On the use of attack graphs for automated planning, we refer the reader to \cite{building}.

\subsubsection{Environment conditions.}
The environment conditions refer
to system configuration or environment properties which may be
necessary or may facilitate the execution of the action.
We distinguish the environment conditions from the requirements,
because the requirements express relations between actions
(which must be taken into account when planning a sequence of actions)
whereas the environment conditions refer to the ``state of the world"
(as far as the attacker is aware of it) before the execution of the module,
and do not imply previous actions. 
For example, an exploit of a buffer overflow that runs only on specific versions
of an operating system, will have as requirement: ``obtain information
about operating system version'' and as environment condition
``OS=RedHat Linux; version between 6.1 and 6.9''. These conditions are
not necessary, as the action can be run anyway, but will dramatically
increase its probability of success.

\subsubsection{Noise produced and stealthiness.}
The execution of the action will produce {\em noise}. This noise
can be network traffic, log lines in Intrusion Detection Systems (IDS), etc. 
Given a list as complete as possible of network sensors,
we quantify the noise produced respective to each of these sensors.
The knowledge of the network configuration and which sensors
are likely to be active, allows us to calculate a global estimate of 
the noise produced by the action. 

With respect to every network sensor, the noise produced can be
classified into three categories: irremovable noise, noise that can
be cleaned if the action is successful (or another subsequent
action is successful), noise that can be cleaned even in
case of failure. So we can also estimate the noise remaining after
cleanup. Of course, the {\em stealthiness} of an action will refer
to the low level of noise produced.

\subsubsection{Running time and probability of success.}
The {\em expected running time} and {\em probability of success}
depends on the nature of the action, but also
on the environment conditions, so their values are updated
every time the attacker receives new information about the environment.
These values are necessary to take decisions and choose a path
in the graph of possible actions. Because of the uncertainties
on the execution environment, we consider three values for the
running time: minimum, average and maximum running time.
Together with the stealthiness and zero-dayness,
these values constitute the cost of the action
and are used to evaluate sequences of actions.

\section{On Vulnerabilities and Exploits}
\label{exploits}

\subsection{Anatomy of an Exploit}

The exploits are the most important actions during an attack.
An exploit is a piece of code that attempts to compromise a workstation or desktop via a specific vulnerability.
According to the literal meaning of exploit, it takes advantage and makes use of a hidden functionality.
When used for actual network attacks, exploits execute payloads of code that can alter, 
destroy or expose information assets. 
When examining an exploit, three main components can be distinguished
(see \cite{payloads} for more information).

\subsubsection{Attack Vector.}
The attack vector is the mechanism the exploit uses to make a vulnerability manifest,
in other words, how to reach and trigger the bug.
For example, in the case of Apache Chunked Encoding Exploit, the attack vector is the 
TCP connectivity that must be established on port 80 to reach the application.
 
\subsubsection{Exploited Vulnerability.}
To obtain an unauthorized result, the exploit makes use of a vulnerability. 
This can be a network configuration vulnerability,
or a software vulnerability: a design flaw or an implementation flaw 
(buffer overflow, format string, race condition).

The most classic example is the buffer overflow, first described in 
``Smashing the stack for fun and profit''  by Aleph One (1996).
The questions for the attacker are: how to insert code and how to modify the execution flow
to execute it?
In the example of a stack based buffer overflow, the code is inserted in a stack buffer
and by overflowing the buffer, the attacker can overwrite the return address and jump to his code.

\subsubsection{Payload.}
Once he manages to trigger and exploit a bug,
the attacker gains control of the vulnerable program.
The payload is the functional component of the exploit, the code the attacker is interested in running.
Classical payloads allow attackers to:

\begin{itemize}
\item{Add a user account: 
on Unix systems, it was done by adding a line to the system password file ({\tt /etc/password})
or changing the password of root.
However such changes are easily detected (Tripwire can detect them)
and to use the account the attacker needs connectivity through legitimate paths (firewalls can block them).
This classical payload is no longer used.
}
\item{Changes to system configuration:
for example, to append a line to the Internet services daemon ({\tt inetd}\footnote{For more information on {\tt inetd},
refer to {\tt http://en.wikipedia.org/wiki/Inetd}})
configuration file on Unix systems ({\tt /etc/inetd.conf})
in order to open a port on the compromised system, 
and later connect to the system via the newly opened port.
}
\item{Shellcode: this is the most popular and has become almost 
synonymous for ``exploit payload".
It comprises opening a shell (a command interpreter), that the attacker can use
to execute available commands.
These payloads are more difficult to detect, but are also more difficult to write.
See the article of Aleph One \cite{alephone} for a recount on this technique.
}
\item{Network aware shellcode.
If the shell is opened on a remote machine, the attacker has to find a way to communicate with the shell.
A first solution is the \emph{bind shellcode} or \emph{bindshell}, 
which listens for incoming connections on a specified network port and protocol (usually TCP).
The problem is that firewalls or other filtering devices may block this connection.
A second option is the \emph{reverse shell}, that initiates the connection from the compromised system.
The third option is the \emph{reuse socket shellcode}, that reuses the connection method used 
to trigger and spawn the shell, thus making use of a communication channel that the attacker knows to work.
}
\end{itemize}

Writing payloads is a very difficult task, that requires solving multiple constraints simultaneously.
The payload is a sequence of byte codes, so each payload will only work in 
a specific operating system and platform.
Depending on the attack vector, the payload may be sent to the vulnerable machine
as an ASCII (American Standard Code for Information Interchange) string (or some protocol field), 
and thus must respect a particular grammar
(for example: byte 0 is forbidden, only 7-bit ASCII is accepted, only alphanumeric characters are accepted, etc.)
Libraries have been developed to help exploit writers to generate shellcodes.
The open source libraries ``MOSDEF"\footnote{The MOSDEF (a short for ``Most Definitely") library is available at\\ 
{\tt http://www.immunitysec.com/resources-freesoftware.shtml }}
and ``InlineEgg"\footnote{The InlineEgg library is available at\\ 
{\tt http://oss.coresecurity.com/projects/inlineegg.html }}
are two well known cases, with tools to cope with the restrictions.
The payload is also typically limited in size (for example the buffer size in the case of a buffer overflow),
so the code that the attacker will run must fit in a few hundred bytes.
 If he wants to execute more complex applications, 
he must find another way.

\subsection{Universal Payload}

We present here the solution called ``syscall proxy'' (developed by Max Caceres and others, see \cite{proxycall} for more details).
The idea is to build a sort of ``Universal Payload'' that allows an attacker
to execute any system call on the vulnerable host.
By installing a small payload (a thin syscall server),
the attacker will be able to execute on his local host complex applications (a fat client),
with all system calls executed remotely.

\subsubsection{Reminder on syscalls.}

A software process usually interacts with certain resources: a file in disk, the screen, a
networking card, a printer, etc. Processes can access these resources 
through \emph{system calls} (\emph{syscalls} for short). These
syscalls are operating system services, usually identified with the lowest layer of communication between a user mode
process and the Operating System (OS) kernel.

Different operating systems implement syscall services differently, sometimes depending on the processor's architecture.
Examples of main groups include UNIX and Windows.

UNIX systems use a generic and homogeneous mechanism for calling system services, usually in the form of 
a ``software interrupt". Syscalls are classified by number and arguments are passed either through the
stack, registers or a mix of both. The number of system services is usually kept to a minimum (about 270 syscalls
can be seen in OpenBSD 2.9), as more complex functionality is provided on higher user-level functions in the libc
library. Usually there's a direct mapping between syscalls and the related libc functions.

In Windows the equivalent functionality is part of large user mode dynamic libraries. We'll refer
to ``Windows syscalls" to any function in any dynamic library available to a user mode process. 
For the sake of simplicity,
this definition includes higher level functions than those defined in {\tt ntdll.dll}, and sometimes very far above the
user / kernel limit.

\subsubsection{Syscall proxy.}

The resources that a process has access to, and the kind of access it has on them, defines the
``context" on which it is executed.
For example, a process that reads data from a file might do so using the open, read and close syscalls.

Syscall proxying inserts two additional layers between the process and the underlying operating system. These layers
are the \emph{syscall client} layer and the \emph{syscall server} layer.
The syscall client layer acts as a link between the running process and the underlying system services. 
This layer is responsible for forwarding each syscall argument and generating a proper request that the syscall server can
understand. It is also responsible for sending this request to the syscall server, usually through the Internet,
 and returning back the results to the calling process.
The syscall server layer receives requests from the syscall client to execute specific syscalls using the underlying
operating system services. This layer marshals back the syscall arguments from the request in a way that the underlying
OS (Operating System) can understand and calls the specific service. After the syscall finishes, its results are marshaled and sent back to
the client, again through the Internet.

There are multiple connection methods between agents. The originating agent can use:
connect to target (similar to bindshell), connect from target (similar to reverse shell), reuse connection and 
Hypertext Transfer Protocol (HTTP) tunneling. 
Agents can also be chained together to reach network resources with limited connectivity.

\begin{figure}[ht]
\centering
\includegraphics[width=12cm]{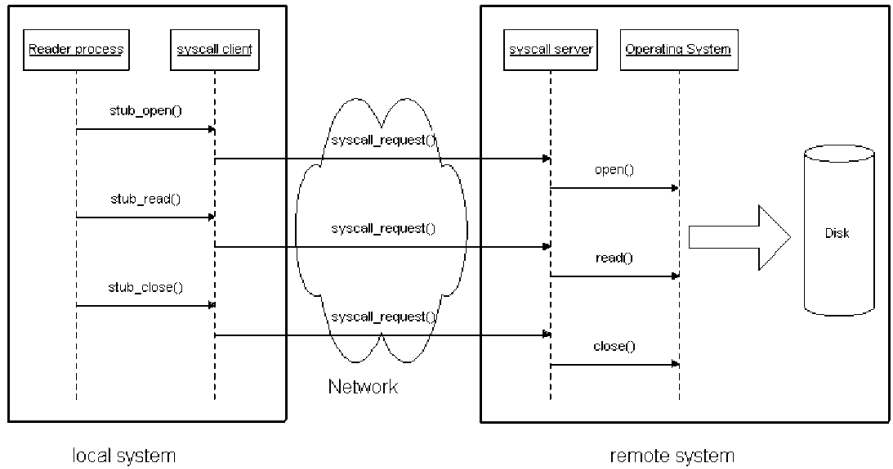}
\caption{Diagram of a Proxy Call Execution}
\label{syscalls1}
\end{figure}

\subsection{Attack Agents}

In our attack model, the abstraction of arbitrary syscall servers
is the concept of Agents.
The Agents are responsible for executing the attack actions.
Thus an attack typically begins with a local agent (representing the attacker, 
which can interact with the local agent through a graphical console),
and follows with the main steps of an attack (as described in \ref{main_steps}),
where the successful exploitation of a vulnerability means installing an agent. 

The attacker is modeled as a set of connected agents, and
exploits are modeled as probabilistic actions
that depend on the details of target OS and applications.

A software agent can take several forms,
examples of which include: scripts, toolkits or other
kinds of programs. The {\em autonomous agents}
who are able to take decisions and continue the attack
without human intervention, are of particular interest.
Such autonomous agents require planning abilities, see \cite{building}

\subsubsection{Agent mission.}

We contemplate different types of organizations between the agents.
One scenario is given by a ``root agent'' who plans the attack and
then gives the other agents orders (of executing actions), eventually
creating new agents if necessary, and asks the agents for 
feedback about action results in order to decide further steps.

Another scenario is when the root agent delegates responsibilities
to the other agents, giving them higher level {\em missions}.
To fulfill the mission, the agent will have to do his own planning
and communicate with other agents. This scenario is likely
to arise when stealthiness is a priority: communications are very
expensive and it becomes necessary to rely on the agents to
execute their missions without giving feedback (or the smallest
amount of feedback, or delayed feedback because of 
intermittent communication channels).

\subsubsection{The environment knowledge.}

The \emph{environment knowledge} (or simply \emph{environment}) is a collection of information
about the computer network being attacked.
Naturally, this information is represented by assets.
In the beginning, the environment contains only 
the {\tt local agent} which will initiate the attack.

The environment plays an important role during the planning phase
and during the execution phase of an attack, since it 
continuously feedbacks the behavior of the agent.
The execution of an \emph{attack action} (as defined in section \ref{actions}) 
makes use of the knowledge that the executing agent has 
of the environment.  
When the action is called, it begins by looking for an asset that
completes its goal in the environment.
If this is the case, the information of the existing asset is used to fulfill the goal, 
and the action returns a success signal, resulting in zero cost 
(in terms of time, noise, success probability and stealthiness).

Note that two interesting graphs can be extracted from the environment knowledge:
the network topology graph and 
the agent distribution graph, whose nodes are the agents involved in the attack
and whose edges are the communication channels between agents.

\section{Large Network Simulator}
\label{simulator}

\subsection{Focus on the attacker's point of view}

We now present our implementation of a network simulator
specifically designed to simulate network attacks.
Our objective is to simulate very large networks,
for example 2.000 machines simulated on a single desktop PC.
It is of course not feasible to simulate all the network traffic,
or to use a VMware server running simultaneously 2.000 images.

The idea of our implementation is to 
focus on the point of view of the attacker.
Using our model of the attacker, we can build a simulator
which is realistic from his point of view.
In particular, the simulator only generates information as requested by the attacker.
By performing this lazy evaluation, the main performance bottleneck comes
from the ability of the attacker to request information from the network.

\subsection{What scenarios can be simulated?}
The simulated scenarios are composed of machines, networking devices and vulnerabilities.
Supported machine components include, but are not limited to: Windows workstations and servers, many Unix
systems, routers, proxies, firewalls and Intrusion Detection Systems (IDS).
Each machine can be independently configured, and installed to run different software
services, such as, but not limited to:
Web server, File Transfer Protocol (FTP) server, Telnet and Secure Shell. 
New applications can be
developed for the simulator platform using the usual development tools.

Network components are used to interconnect machines, and can simulate hubs,
switches, vlans and dialup connections and their security characteristics.
Vulnerability descriptions are entered in the vulnerabilities database, to allow the
simulation of the vulnerable application behavior. (There is no need to modify the actual
application's code to reflect a vulnerability, nor to write any exploit code). Aspects and
types of vulnerabilities simulated include, but are not limited to: local / remote, denial of service / exploit /
leakage, probabilistic / dependent on hidden parameters and noise level.

\subsection{Multiplatform Agents}

According to our attack model, an attacker can be effectively modeled by a set of agents.
Thus by simulating the behavior of the agents,
we can simulate the behavior of the network (this is transparent for the attacker).
The whole environment is accessed by the attacker through the local agent, 
and interactions take place in the form of proxied system calls.

The base of the simulator are agents that respond to proxycalls.
The agents implement a syscall server for a specific operating system and platform, 
but all of them have the same interface: this is what we call ``Multiplatform Syscall". 
So, if an attacker (client) can install a multiplatform syscall agent in a victim host, 
he does not care about the syscalls supported by the target host, 
the attacker merely needs to know the universal syscall interface exported by the agent.

\subsection{The Semantics of the Exploit Database}

\subsubsection{Security Model.}
In the simulator security model, a \emph{vulnerability} is a mechanism used to (potentially) access an otherwise 
restricted communication channel.
An \emph{exploit} is a ``magic" string that opens access to some vulnerable agent's channel.
It can be simulated as a message with a symbolic identifier, sent to an application.
Depending on the environment conditions, the exploit database will determine the resulting behavior
of the application.

Given a target machine $M$, the simulator iterates the list-like structure of \emph{results} in order. 
Each result entry has conditions associated  to it, so the simulator iterates the tree-like structure of  \emph{requirements}
section and, if a match is found, the action (install an agent, crash or reset) is executed with probabilistic behavior. 
The  execution of actions stops when an action is evaluated to True.
 
\subsubsection{Requirements.}

In the \emph{requirements} section, you can use several kind of  tags. 
The tags specify the conditions that have influence on the execution of the exploit 
(that is on the result probabilities). 
Example:
\begin{verbatim}
<requirement type="system" id="req0">
  <os arch="i386" name="windows" /> 
  <win>nt4</win> 
  <edition>server enterprise_server</edition> 
  <servicepack>6 6a</servicepack> 
</requirement>
\end{verbatim}
This states that one of the possibilities is that the target machine runs  Windows version NT4, 
the edition should be ``server" or ``enterprise\_server" and  the service pack should be 6 or 6a. 
The requirements have a unique id to  identify them, in this case ``req0".
Another requirement concerns the target application:
\begin{verbatim}
<requirement type="application" id="req1">
  <status>target</status> 
  <name>Internet Information Services</name> 
  <version major="4 5" /> 
</requirement>
\end{verbatim}
This states that the machine should be running Internet Information  Services (IIS), version major 4 or 5, 
and this application is the target of the  exploit.

The possible status are: \\
1. \emph{target}: the application is the target of the exploit (the most common  case). \\
2. \emph{running}: the application should be running but is not necessarily the  target of the attack. \\
3. \emph{installed}: the application should be installed but not necessarily  running. \\
4. \emph{not running}: for  example, a remote exploit will have more success probability if the target  machine is in a network 
with no firewalls running.

Requirements can be combined, for example:
\begin{verbatim}
<requirement type="compose" id="req2">
  <operator>logic_and</operator> 
  <operands>req0 req1</operands> 
</requirement>
\end{verbatim}
The result of the ``logic\_and" operation is a requirement stating that the  target machine should be running 
Windows NT4 server edition or  enterprise\_server edition, and running IIS (Internet Information Services). 
 There is also a ``logic\_or".

\subsubsection{Results.}

The result is a list of the relevant probabilities, for example:
\begin{verbatim}
<result for="req1">
  <crash chance="0.00" what="os" /> 
  <reset chance="0.00" what="os" /> 
  <crash chance="0.10" what="application" /> 
  <reset chance="0.00" what="application" /> 
  <agent chance="0.75" /> 
</result>
\end{verbatim}

In order, these are: the chance of crashing the machine, 
of resetting the  machine (reboot), 
of crashing the target application (IIS), 
of resetting the  target application, and of successfully installing an agent.

To determine the result, we follow this procedure: 
processing the lines in  order; and for each positive probability, 
choose a random value between 0 and 1.  If the value is smaller than the chance attribute, 
the corresponding action is  the result of the exploit. 

In this example, we draw a random number to see if the application crashes. 
If  the value is smaller that 0.10, the application Internet Information Services (IIS) is crashed 
and the  execution of the exploit is finished. 
Otherwise, we draw a second number to  see if an agent is installed. 
If the value is smaller than 0.75, the agent is  installed, otherwise there is no visible result.

Other possible results (to be implemented) are: \\
1. Raise an Intrusion Detection System (IDS) alarm. \\
2. Write some log in a network actor (like a firewall, IDS, router, etc).  \\
3. Capture a session id, cookie, credential or password.

\section{Performance Issues} \label{performance}


\subsection{Simulation versus emulation}

From a systemic view-point, 
we speak of simulation when the level of detail of interaction between components 
inside the system is mimicked, and emulation when only the interaction of the system 
with the environment is mimicked. 
These definitions depend on the level of abstraction, 
on a level of abstraction we can describe the behavior of the system as a black-box 
or we can describe the behavior of the components individually.

Following this line of thought, the system implemented simulates networks in the socket abstraction level, 
and inside the network the behavior of machines is emulated from the communication angle.
The emulation of computers is basic but complete, in the sense that a remote virtual user  
connecting to one of them can execute different processes and handle data files.

\subsection{Tension between realism and performance}
 
 There is a tension between realism and performance in the simulation. 
 In this case, good performance is achieved by only simulating the syscalls / socket abstraction level. 
Most actions work at the syscall level and attack upper levels of abstraction, 
whereas the network packet switching is not simulated.

The network simulator was designed to be able to simulate networks of thousands of computers. 
Each simulated machine has at least one thread. 
The goal was to have a simulator on a single desktop computer with a simulated traffic realistic 
from a penetration test point of view. It was not designed to simulate Distributed Denial of Service (DDoS) attacks as floods or worms 
but there is a possibility in that direction also, maybe in dedicated servers running the simulation.

\subsection{Socket direct}

A hierarchy for file descriptors was developed, including a variety of sockets optimized for the simulation 
in one computer, called ``socketdirect".
Socket direct is fast:  as soon as a connection is established, the client keeps a filedescriptor
pointing directly to the server's descriptor. Routing is only executed during the connection.
Process Control Blocks (PCB) are created as expected, but are only used during connection establishment.
There is support for 
Transmission Control Protocol (TCP) and 
User Datagram Protocol (UDP) sockets, and a central set of systems calls, 
including filesystem syscalls, to emulate memory in each machine of the network.
 Data enters to the simulation through the socket subclass ``socketreal", 
which wraps a real 
BSD (Berkeley Software Distribution) socket of the underlying operating system.

\subsection{Scheduler}

The responsibility of the scheduler is to assign the 
Central Processing Unit (CPU) resources to the different machines
in the simulation and inside each machine to the different processes.
The scheduling is non preemptive and round-robin. 
The scheduling iterates over the hierarchy machine / process / thread as a tree
(like a depth-first search). 
Each machine has the possibility to run in round-robin,
 where running means that the machine runs its processes in round-robin.
 The same way running a process is giving all its threads the order to run until a syscall is needed, if possible. 
 Obviously depending on the state of each thread, they run, change state or finish execution. 
The central issue is that threads execute systems calls and then if possible continue their activity until they finish
 or another system call is required. 

Simulated threads are real threads of the OS (Operating System), simulated machines and
 processes are all running within the unique process of the simulator.
Thanks to this architecture, there is no loss of performance due to context switching
(descriptors and pointers remain valid when switching from one machine to the other).

Something to remark is that the simulator doesn't have to use all the CPU when idle, so the scheduler was
 devised to sleep (e.g., 20ms) the simulator after executing a constant number (e.g., 512) of machine runs (runs to sleep). 
This leaves space for other programs interacting with the simulator to continue their normal activity in the desktop machine
while the simulator is idle.

 Another improvement was to change the runs to sleep dynamically in a exponential increment and linear back-off fashion,
 depending on a threshold of syscalls lost per sleep. This results in better overall response when there is simulated 
activity and less use of the CPU when there is no simulation activity. 

\begin{figure}[t]
\centering
\includegraphics[width=12cm]{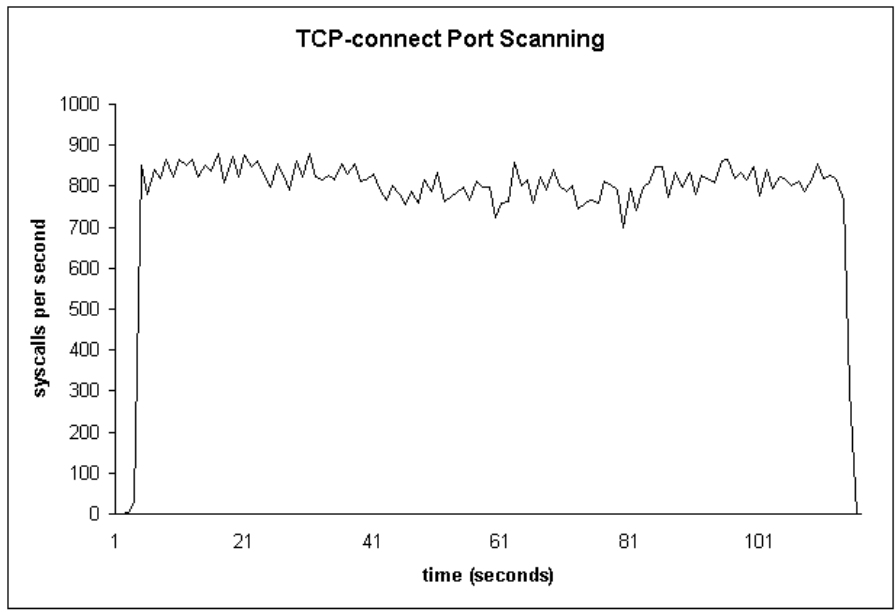}
\caption{Simulated TCP Port Scanning (time versus syscalls/second)}
\label{fig:syscalls1}
\end{figure}

\begin{figure}[t]
\centering
\includegraphics[width=12cm]{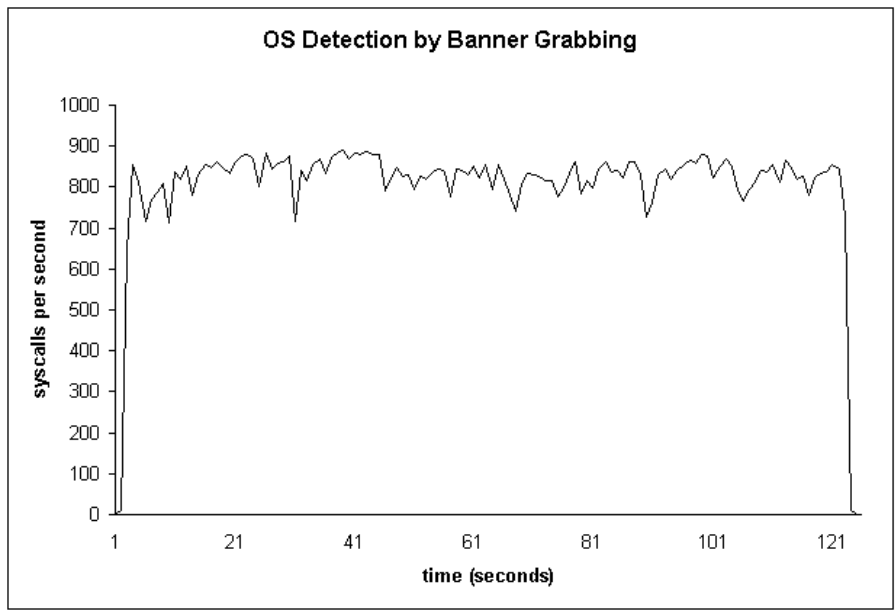}
\caption{Simulated OS Detection by Banner Grabber (time versus syscalls/second)}
\label{fig:syscalls2}
\end{figure}

Figures \ref{fig:syscalls1} and \ref{fig:syscalls2} show measurements that were done on a 
Pentium D 2.66Ghz machine with 1.5GB of RAM, running Windows XP SP2.
The simulated scenario includes 100 networks of 10 machines each, 
so there are 1.000 machines running in the simulation.
When responding to a TCP Port Scan or an OS Detection by Banner Grabber, 
the simulator answers between 700 and 900 syscalls per second.
Total running time of the modules (on a single network) lies between 100 and 120 seconds.

\subsection{File system with templates}

In order to handle thousand of files, avoiding wasting of huge disk space, 
the filesystem simulation is accomplished 
mounting template filesystems. 
A \emph{template filesystem} is a common file repository shared by a group of virtual machines. 
For example, all Windows systems share a common Windows file repository with the default installation provided by Microsoft. 
These templates have read permission only, so when a machine needs to read or change a file, 
the file is copied to a local filesystem in that machine, this technique is well know as ``copy on write".
 The fundamental idea is that if multiple callers ask for resources that are initially indistinguishable, 
you can give them pointers to the same resource. This function can be maintained until a caller tries to modify
 its copy of the resource, at which point a true private copy is created to prevent the changes 
from becoming visible 
to everyone else. All of this happens transparently to the callers. The primary advantage is that if a caller never 
makes any modification, no private copy needs to be created.

In order to improve the performance, a file cache was implemented: the simulator saves the most recent accessed files
 (or block of files) in memory. In high scale simulated scenarios, it is very common to have several machines 
doing the same task at (almost) the same time. For example, when the system starts up, 
all UNIX machines read the boot script from {\tt /etc/initd} file; 
if these kinds of files are in the system cache,
 the booting process is faster, because only few disk accesses are needed, 
even in scenarios of hundreds or thousands of simulated machines.

\section{Conclusion}

We presented a network simulator focused on the attacker's point of view.
The simulation is based on a model of network attacks, whose building blocks
are Assets, Actions and Agents.
By making use of the proxy syscalls technology, and simulating multiplatform agents,
we were able to implement a simulation that is both realistic and 
light-weight, allowing the simulation of networks with thousands of hosts.
Some applications of the simulator are:
\begin{itemize}
\item{Cyber attack modeling and analysis tool.
The different security components can be configured to report attack evidence
in the same way as the real world components, allowing, for example, post-attack
forensics analysis and real-time detection exercises.
}
\item{Pentest training tool.
A step by step tutorial for pentesters is hard to write because
the user might not have a proper target network setting,
or because the characteristics of the user's target network are unknown.
The simulator can be used to deploy several complex scenarios in the user's computer, 
so that the user can follow the training on a shared scenario.
}
\item{Evaluation of countermeasures.
Consider a system administrator that has a set of measures
which make certain attack actions less effective (in our framework, a measure
may reduce the probability of success of an attack action, or increase
the noise it produces, for example by adding a new IDS).
He can then use the simulation to see if his system becomes safe
after all the measures are deployed, or to find a minimal set
of measures that make his system safe.
}

\end{itemize}



\end{document}